\documentstyle[prl,aps]{revtex}
\begin{document}

Pis'ma v ZhETF 66 (1997) 275-279

\centerline{\large \bf Single-electron computing without dissipation}

\vskip 2mm

\centerline{A.M.Bychkov, L.A.Openov $^{a)}$, I.A.Semenihin}

\vskip 2mm

\centerline{\it Moscow State Engineering Physics Institute
(Technical University)}
\centerline{\it 115409 Moscow, Russia}

\vskip 4mm

A possibility to perform single-electron computing without dissipation
in the array of tunnel-coupled quantum dots is studied theoretically,
taking the spin gate NOT (inverter) as an example. It is shown that the
logical operation can be realized at the stage of unitary evolution of
electron subsystem, though complete switching of the inverter cannot be
achieved in a reasonable time at realistic values of model parameters. An
optimal input magnetic field is found as a function of inter-dot
tunneling energy and intra-dot Coulomb repulsion energy.

\vskip 8mm

Recent advances in the fabrication of nanometer scale quantum dots open an
opportunity of practical implementation of the idea to use the states of a
quantum system for data coding and processing \cite {Feynman}. For example,
the spins of individual electrons can be viewed as the bits of information:
logical one (zero) corresponds to "up" ("down") direction of
electron spin at a given quantum dot. {\it Spin gates} (elementary sets of
quantum dots performing particular logical functions) have been discussed by
Bandyopadhyay {\it et al} \cite {Bandyo} and later investigated
theoretically by Molotkov and Nazin \cite {Molotkov} and by Krasheninnikov
and Openov \cite {Openov}. If occupation/unoccupation of a quantum dot by a
single electron is viewed as a bit 1/0, one has {\it charge gates}, various
kinds of which have been studied, e.g., by Lent {\it et al} \cite {Lent} and
by Nomoto {\it et al} \cite {Nomoto}. In arrays of quantum dots, the quantum
tunneling of electrons between adjacent dots and/or Coulomb interaction of
electrons with each other play a role of "wiring", resulting in the signal
propagation from dot to dot.

The operation of spin and charge gates is based on the principle of "ground
state computation" \cite {Bandyo},\cite {Lent}. According to this
principle, upon the influence of external source on input dots of a
particular gate, the electron subsystem changes to a new ground state. The
final spin or charge configuration reflects the result of "calculation".
This result can be read from output dots of the gate. The quantum dot gates
are believed to possess high-speed performance as a consequence of extremely
fast switching between different electron ground states. However, the
switching rate, being dictated by dissipation processes, is not known {\it a
priori}. For a gate consisting of a few quantum dots and operating at
sufficiently low temperatures, the switching rate may appear to be rather
small ($10^{6}$ to $10^{9}$ s$^{-1}$ \cite {Nomoto}), thus slowing down the
computation. Hence, in studies of the potential of quantum dot arrays for
high-speed single-electron computing one should give special attention to
inelastic relaxation processes.

An alternative way has been discussed recently by Bandyopadhyay and
Roychowdhury \cite {Bandyo2}. They explored the dynamic behavior of the
simplest spin gate NOT (inverter) and found that there exists optimal
input signal energy to achieve its {\it complete} switching in the absence
of inelastic relaxation. However, it remains unclear how adequately the
results obtained in \cite {Bandyo2} depict reality since the authors of
Ref.\cite {Bandyo2} used the Heisenberg model to describe the correlated
electrons in quantum dots. Meanwhile, it is well known that this model is
just a limiting case of more realistic Hubbard model \cite {Izyumov} and
cannot be used if the inter-dot electron tunneling energy $V$ is of the
order or greater than the intra-dot electron repulsion energy $U$. It is
instructive to study the broad range of $V$ and $U$ values in order to
see if the conclusions of \cite {Bandyo2} reflect the basic physics or
are just a consequence of using a particular theoretical model.

In this paper we study the unitary evolution of electron subsystem in the
spin gate NOT (inverter) making use of the Hubbard model with arbitrary
values of $V$ and $U$. The inverter consists of two closely spaced quantum
dots (A and B) occupied by two electrons \cite {Bandyo},\cite
{Molotkov},\cite {Bandyo2}. One of two dots (say, the dot A) serves for
writing the input signal to the gate by the action of the local magnetic
field $H_A$. The second dot (B) is the output. At $H_A = 0$ the ground state
of the inverter is the entangled state with zero magnetic moments at both
dots A and B. The logical function NOT is realized if at $H_A \neq 0$
magnetizations (i.e., spin projections) of dots A and B have opposite
directions. {\it Complete} switching of the inverter is said to take place
if spin projections are saturated ($S_{zA}=1/2$, $S_{zB}=-1/2$ or
$S_{zB}=1/2$, $S_{zA}=-1/2$, where $S_{zi}=\langle\hat{S}_{zi}\rangle=
\langle\hat{n}_{i\uparrow}-\hat{n}_{i\downarrow}\rangle/2$, $\hat{n}_i$
being the operators of particles number at dots $i=A,B$). Upon complete
switching, magnetizations of both dots reach the maximum absolute value
$g\mu_B$, where $g$ is the Lande factor, $\mu_B$ is the Bohr magneton. We
stress that the ground state at any {\it finite} value of $H_A$ is organized
in such a way that $|S_{zA}|<1/2$ and $|S_{zB}|<1/2$ \cite {Molotkov},\cite
{Bandyo2}.  Hence, the complete switching of the inverter cannot be achieved
through its {\it inelastic relaxation} to a new ground state.

The Hubbard Hamiltonian for the inverter has the form
\begin{equation}
\hat{H}=-V\sum_{\sigma}
(\hat{a}^{+}_{A\sigma}\hat{a}^{}_{B\sigma} +
\hat{a}^{+}_{B\sigma}\hat{a}^{}_{A\sigma})+
U\hat{n}_{A\uparrow}\hat{n}_{A\downarrow}+
U\hat{n}_{B\uparrow}\hat{n}_{B\downarrow}-
g\mu_B H_A \sum_{\sigma}\hat{n}_{A\sigma}sign(\sigma),
\end{equation}
where the quantities $V$, $U$, $H_A$, $g$, and $\mu_B$
are defined above, other notations being standard (see, e.g., \cite
{Openov}). Here we assume that each dot has one size-quantized level with
on-site potential $\varepsilon_0=0$ (i.e., all energies are measured from
$\varepsilon_0$). In what follows, we set $g\mu_B=1$.

The complete orthonormal set of inverter eigenstates is formed by
two-electron basis states
$|1\rangle$=$|\uparrow,\downarrow\rangle$,
$|2\rangle$=$|\downarrow,\uparrow\rangle$,
$|3\rangle$=$|\uparrow\downarrow,0\rangle$,
$|4\rangle$=$|0,\uparrow\downarrow\rangle$,
$|5\rangle$=$|\uparrow,\uparrow\rangle$,
$|6\rangle$=$|\downarrow,\downarrow\rangle$,
where, e.g., the notation $|\uparrow,\downarrow\rangle$ denotes the state
with up-spin electron at the dot A and down-spin electron at the dot B, the
notation $|\uparrow\downarrow,0\rangle$ denotes the state with two (up-spin
and down-spin) electrons at the dot A and no electrons at the dot B, {\it etc}.
The magnetic moment of the electron with up-spin polarization is oriented
along the direction of the local applied magnetic field $H_A$.

At $H_A=0$, the ground state eigenvector of the Hamiltonian (1) is
\begin{equation}
\Psi_{0}=\frac{1}{2}\sqrt{1+\frac{U}{\sqrt{U^2+16V^2}}}\left(|1\rangle+
|2\rangle+\frac{\sqrt{U^2+16V^2}-U}{4V}|3\rangle+
\frac{\sqrt{U^2+16V^2}-U}{4V}|4\rangle\right).
\end{equation}
The corresponding eigenenergy is $E_0=(U-\sqrt{U^2+16V^2})/2$.
In the ground state we have $\langle\Psi_0|\hat{S}_{zA}|\Psi_0\rangle
=\langle\Psi_0|\hat{S}_{zB}|\Psi_0\rangle=0$. We suppose that at $t\le 0$
the system is in its ground state.

If the local external magnetic field is applied at time $t=0$, then the
wave function $\Psi(t)$ at $t\ge0$ is
\begin{equation}
\Psi(t)=\sum_{k=1}^{6}A_k\Psi_k\exp(-iE_kt/\hbar),
\end{equation}
where $\Psi_k$ and $E_k$ $(k=1-6)$ are eigenvectors and eigenenergies of the
stationary Schr\"odinger  equation
\begin{equation}
\hat{H}\Psi_k=E_k\Psi_k.
\end{equation}
The coefficients $A_k$ should be found from the initial condition
$\Psi(t=0)=\Psi_0$. It is convenient to write $\Psi_k$ as
\begin{equation}
\Psi_k=\sum_{n=1}^{6}B_{kn}|n\rangle.
\end{equation}
Then
\begin{equation}
\Psi(t)=\sum_{n=1}^{6}f_n(t)|n\rangle,
\end{equation}
where
\begin{equation}
f_n(t)=\sum_{k=1}^{6}A_kB_{kn}\exp(-iE_kt/\hbar).
\end{equation}
The probability to find the system in the
basis state $|n\rangle$ at time $t$ is $p_n(t)=|f_n(t)|^2$.

At arbitrary values of $V$, $U$, and $H_A$ the eigenvalue equation (4)
reduces to the algebraic equation of the third power in $E_k$. The resulting
analytic expressions are too cumbersome for analysis, so it is more
convenient to solve the equation (4) numerically. Before proceeding to the
results of these calculations, let us consider the limiting case $U=0$ which
physically corresponds to $U<<V$ (closely-spaced large-sized dots \cite
{Nomoto}).

At $U=0$ we have rather simple equations for the probabilities $p_n(t)$:
\begin{eqnarray}
&&p_1(t)=\frac{1}{4}\left(1+\frac{4H_AV}{H^2+4V^2}\sin^2(\omega t/2)\right)^2 ,~
p_2(t)=\frac{1}{4}\left(1-\frac{4H_AV}{H^2+4V^2}\sin^2(\omega t/2)\right)^2 ,\nonumber \\
&&p_3(t)=p_4(t)=\frac{1}{4}\left(1-\frac{16H^2_AV^2}{(H^2+4V^2)^2}\sin^4(\omega t/2)\right) ,~
p_5(t)=p_6(t)=0 ,
\end{eqnarray}
where $\omega=\sqrt{H^2_A+4V^2}/\hbar$. From (6) and (8) it is
straightforward to find that
\begin{equation}
S_{zA}(t)=\langle\Psi(t)|\hat{S}_{zA}|\Psi(t)\rangle=
-S_{zB}(t)=(p_1(t)-p_2(t))/2=\frac{2H_AV}{H^2_A+4V^2}\sin^2(\omega t/2).
\end{equation}

From (9) we see that the spins $S_{zA}$ and $S_{zB}$ are oppositely directed
at any time $t$ according to the physical truth table of the logical gate NOT
\cite {Molotkov}. For the sake of definiteness, let us consider the case
$H_A>0$. In this case $S_{zA}$ is always positive and peaks at
$t_0=\pi/\omega$. Moreover, a {\it complete} switching, $S_{zA}(t_0)=1/2$
and $S_{zB}(t_0)=-1/2$, is achieved at $H_A/V=2$. The dependence of
$S_{zA}(t_0)$ on $H_A/V$ is shown in Fig.1. We stress that this dependence
has been obtained by us in the weak coupling limit of the Hubbard model.
Nevertheless, it is analogous to those calculated in \cite
{Bandyo2} within the Heisenberg model (i.e., in the strong coupling limit of
the Hubbard model, $U>>V$) with an exception that in the Heisenberg model
$\omega=\sqrt{H^2_A+4J^2}/\hbar$, $t_0=\pi/2\omega$, and $S_{zA}(t_0)$
reaches a maximum value of 1/2 at $H_A=2J$ \cite {Bandyo2}, where J is the
antiferromagnetic exchange energy (for two-site cluster $J=V^2/U$ at
$U>>V$). Hence, one may expect that complete switching of the inverter can
occur at an {\it arbitrary} ratio of $U$ to $V$.

To check this hypothesis, we calculated numerically the dependencies of
$S_{zA}$ on $t$ and $S_{zA}(t_0)$ on $H_A/V$ at different values of $U/V$,
where $t_0$ is generally defined as a time of the first maximum at the curve
$S_{zA}(t)$, $t_0$ being a function of $H_A/V$ and $U/V$. The curves of
$S_{zA}(t_0)$ versus $H_A/V$ are shown in Fig.1 for several values of $U/V$.
One can see that increase in $U/V$ first results in decreased height of the
maximum at the $S_{zA}(t_0)$ versus $H_A/V$ curve. At $U/V>2$ the height of
this maximum increases again, but doesn't reach the saturated value 1/2 at
finite $U/V$, though $S_{zA}(t_0)\rightarrow 1/2$ if $U/V\rightarrow\infty$
(this corresponds to the Heisenberg model and agrees with the results
obtained in \cite {Bandyo2}).

It seems that complete switching of the inverter cannot be achieved at
realistic values of $U/V$ ratio, i.e., at $U/V\neq 0$ and $U/V\neq \infty$.
Note, however, that at arbitrary values of $H_A/V$ and $U/V$ the function
$S_{zA}(t)$ is not periodic in time since it includes several harmonics
with different frequencies and amplitudes. Hence, in principle, the value
of $S_{zA}=1/2$ can be achieved at some longer time. But this case is of no
interest for us since we should like not only to reach the maximum
permissible value of $S_{zA}$, but to do it in as short as possible
switching time.

However, an impossibility to achieve the complete switching of the inverter
doesn't imply the impossibility to perform the logical operation NOT at
the stage of the unitary evolution. One should just to "read" the signal at
a time when $S_{zA(B)}$ has a large absolute value, e.g., at a time $t_0$.
Indeed, $S_{zA}(t_0)\ge 0.45$ at any value of $U/V$ (see Fig.1). Hence, the
error probability $P_{err}=1-p_1(t_0)$ (i.e., the probability to read the
"wrong" signal $S_{zA}=-1/2$ or $S_{zA}=0$ at a time $t_0$) is less than 0.1.
Our calculations showed
that at $U/V<<1$ and at 'optimal' (for a given $U/V$) value of $H_A/V$ the
time $t_0$ is of the order of $\hbar/V$, i.e., $t_0\approx 10^{-13}s$ for
$V\approx 10$ meV. The value of $t_0$ increases as $U$ increases
and reaches $\approx 6\hbar/V$ at $U/V=10$. The limiting value of $t_0$
at $U>>V$ is $t_0=\pi\hbar U/4\sqrt{2}V^2$, in accordance with
\cite {Bandyo2}. Thus, to speed up the calculation, we should have small
$U$ and large $V$. If the shape of a single quantum dot is a cube with
side length $a$ and the distance between the quantum dots is $d$, then
$V$ decreases exponentially in both $d$ and $a$, while $U$ is roughly
inverse proportional to $a$ and is almost independent on $d$
\cite {Nomoto},\cite {Openov2}. Hence, small values of $d$ and $a$
favor short switching times $t_0$ (the values of $U$ and $V$ can be
calculated numerically for a given set of geometrical parameters of
quantum dots array and for a particular choice of semiconducting materials
\cite {Nomoto}).

On the other hand, the 'optimal' value of $H_A$ increases with $V$ as
$H^{opt}_A=2V$ at $U=0$ (9) and as $H^{opt}_A=2V^2/U$ at $U/V>>1$ (see also
\cite {Bandyo2}). The product $H^{opt}_At_0$ is of the order of $\hbar$
at any $U/V$. Hence, one can have a realistic value of $H^{opt}_A<1$ Tesla
only at the expense of increasing $t_0$ up to $\approx 10^{-11}$s.
Nevertheless, this value of $t_0$ still remains several orders of magnitude
smaller than characteristic times of inelastic relaxation \cite {Nomoto}.

In summary, the switching of the spin gate NOT (inverter) at the stage of
unitary evolution is much faster than through relaxation to a new ground
state. The switching time can be reduced down to $10^{-11}$s through proper
choice of quantum dots geometrical parameters and local external magnetic
field, with the error probability less than 0.1.

This work was supported by the Russian Foundation for Fundamental Research
under Grant No 96-02-18918. We are grateful to S.Bandyopadhyay for sending
us a preprint of the work \cite {Bandyo2} prior to publication. We thank
S.N.Molotkov for numerous valuable comments. Useful discussions with
V.F.Elesin, A.V.Krasheninnikov and S.S.Nazin are gratefully acknowledged.

\vskip 4mm

$^{a)}$L.A.Openov's e-mail: opn@supercon.mephi.ru

\vskip 6mm

\centerline{\bf Figure caption}
\vskip 2mm

Fig.1. The maximum value of $S_{zA}(t_0)$ versus $H_A/V$ at different $U/V$.


\begin{references}

\bibitem{Feynman}R.Feynman, Uspekhi Fiz. Nauk {\bf 149}, 671 (1986).
\bibitem{Bandyo}S.Bandyopadhyay, B.Das, and A.E.Miller, Nanotechnology
{\bf 5}, 113 (1994); S.Bandyopadhyay, V.P.Roychowdhury, and X.Wang,
Phys. Low-Dim. Struct. {\bf 8/9}, 28 (1995).
\bibitem{Molotkov}S.N.Molotkov and S.S.Nazin, Pis'ma v ZhETF {\bf 62}, 256
(1995); ZhETF {\bf 110}, 1439 (1996).
\bibitem{Openov}A.V.Krasheninnikov and L.A.Openov, Pis'ma v ZhETF {\bf 64},
214 (1996).
\bibitem{Lent}C.S.Lent and P.D.Tougaw, J. Appl. Phys. {\bf 74}, 6227 (1993).
\bibitem{Nomoto}K.Nomoto, R.Ugajin, T.Suzuki, and I.Hase, J. Appl. Phys.
{\bf 79}, 291 (1996).
\bibitem{Bandyo2}S.Bandyopadhyay and V.P.Roychowdhury, Superlattices and
Microstructures, to be published.
\bibitem{Izyumov}Yu.A.Izyumov, Uspekhi Fiz. Nauk {\bf 165}, 403 (1995);
Uspekhi Fiz. Nauk {\bf 167}, 465 (1997).
\bibitem{Openov2}A.V.Krasheninnikov, S.N.Molotkov, S.S.Nazin, and
L.A.Openov, ZhETF {\bf 112} (1997), to be published.

\end{references}
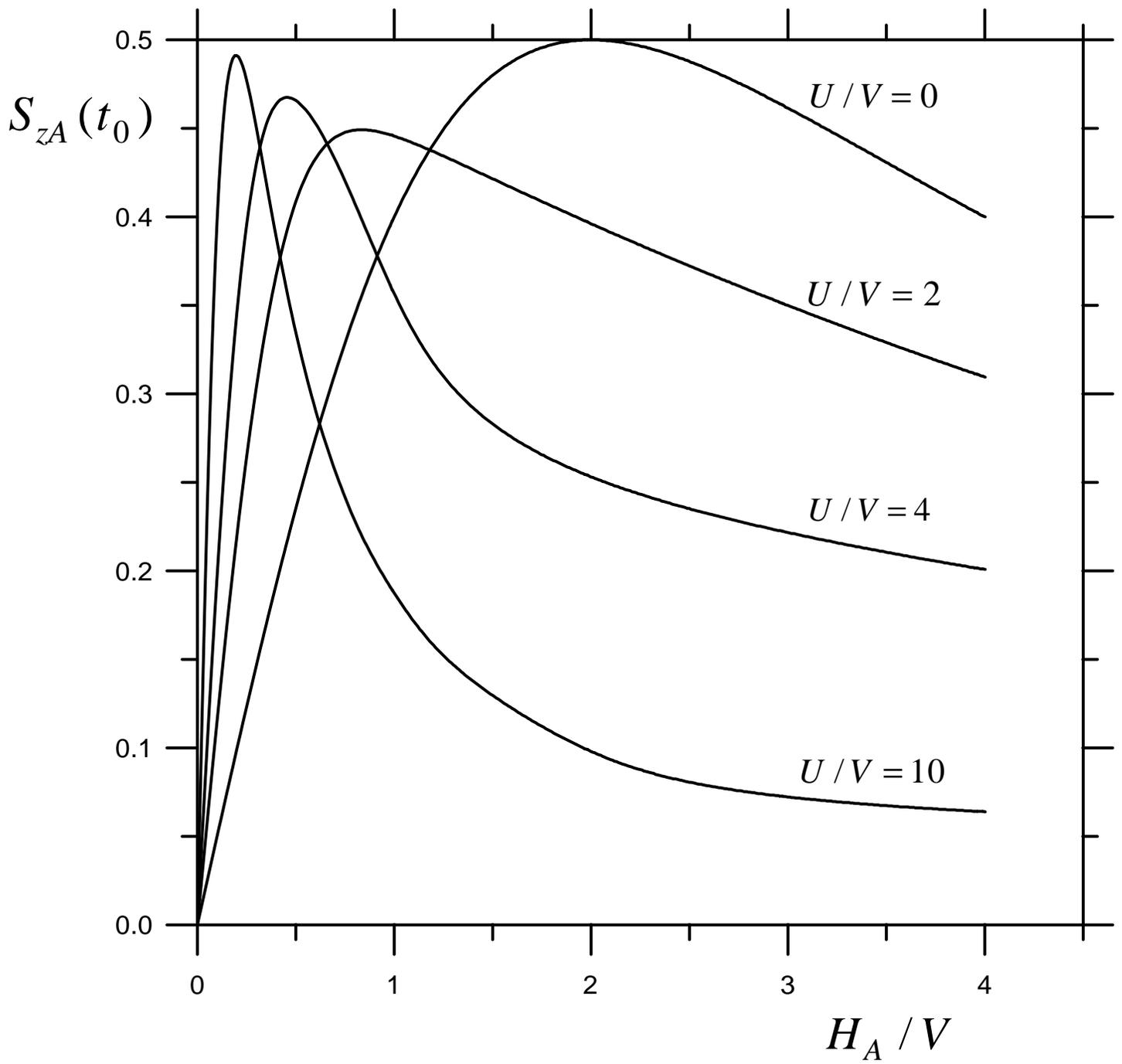
\end{document}